\documentclass[12pt]{iopart}
\usepackage{iopams,graphicx}

\usepackage{lmodern}
\usepackage[T1]{fontenc}
\usepackage{bbold}
\graphicspath{{Graphics/}}
\usepackage{xcolor}

\usepackage{soul}
\usepackage[mathscr]{euscript}
\usepackage{cite}
\usepackage[super]{nth}
\usepackage{float}  
\usepackage{comment}   
\usepackage[colorlinks, citecolor = blue, urlcolor = blue]{hyperref}

\begin{document}

\title[Sluggish random walker with memory]{Ultra slow sub-logarithmic diffusion of a sluggish random walker subject to resetting with memory}
\author{Denis Boyer}
\address{Instituto de F\'\i sica, Universidad Nacional Aut\'onoma de M\'exico, 
Ciudad de M\'exico 04510, M\'exico}
\author{Satya N. Majumdar}
\address{LPTMS, CNRS, Univ.  Paris-Sud,  Universit\'e Paris-Saclay,  91405 Orsay,  France}

\begin{abstract}

We solve a model of sluggish stochastic motion in which a Brownian particle diffuses with a diffusion coefficient that decays algebraically with the distance to the 
origin, as $|x|^{-\alpha}$. Additionally, the particle resets with a constant rate $r$ to 
positions previously visited in the past, so that frequently visited regions are more likely 
to be revisited. An exact expression is obtained at all times for the position distribution 
in arbitrary spatial dimensions. At late times, the typical displacement of the walker from 
the origin grows extremely slowly, as $[\ln(rt)]^{1/(\alpha +2)}$, and the 
position distribution tends to a scaling law. For any $\alpha>0$, the scaling function 
has a bimodal shape with a minimum at $x=0$ and non-Gaussian tails. Although the mean 
square displacement is hard to compute, some generalized moments of this 
process can be calculated exactly at all times in one dimension, and are shown to be closely 
related to the moments of the well-studied model with a constant diffusion coefficient.

\end{abstract}

\maketitle

\section{Introduction}

Resetting with memory has been demonstrated to slow down the dynamics of a diffusing particle
considerably. Consider, for instance, a single Brownian particle on a line
with a diffusion constant $D$. In addition 
to diffusion, the particle undergoes a resetting process with rate $r$ 
to previously visited positions. 
Resetting follows a stochastic rule that incorporates memory effects: the particle that 
resets at some time $t$ chooses uniformly at random a preceding time $0\le t'\le t$, i.e., with probability density $1/t$, and occupies the position at which it was located at 
$t'$ again. This resetting move with memory favours revisit to frequently visited locations and therefore tends
to localize the particle, while the normal diffusion tends to de-localize it.
Eventually de-localization wins but at the cost of a considerable slowing down
of the dynamics. The position distribution of this particle at time $t$ was first computed exactly
on a lattice~\cite{BS2014}, and later in a continuous space-time version of the model~\cite{BEM2017}.
At late times, the position distribution $P_r(x,t)$ turns out to be a simple Gaussian, albeit with
a variance that grows logarithmically with increasing time~\cite{BS2014,BEM2017}
\begin{equation}
P_r(x,t) \approx \sqrt{ \frac{r}{4\, \pi\, D\, \ln(rt)}}\, \exp\left[-\frac{r\, x^2}{4\, 
D\, \ln (rt)}\right]\, .
\label{diff_mem.1}
\end{equation}
This slow dynamics can be traced back to the fact that 
the resetting induced memory makes the walker likely to re-visit frequently visited locations and thus growth via visiting fresh sites gets severely hindered.

This simple model of diffusion with resetting to a previously visited location 
via a uniform memory kernel has been generalized in various directions.
This includes incorporating a fading memory~\cite{BR2014}, generalization to L\'evy flights~\cite{BP2016},
generalization to a broad memory kernel going beyond the uniform case~\cite{BEM2017} and more
recently to active run and tumble particles~\cite{BM2024}. A Brownian particle in one dimension in the presence of a 
uniform resetting induced memory and an external confining potential $U(x)$ has been exactly solved recently~\cite{BM2024_1}.
This study demonstrated that, somewhat unexpectedly, the system reaches a Gibbs-Boltzmann stationary state
$P^{\rm st}(x)\propto e^{-U(x)/D}$ at long times, but the relaxation to this stationary state is rather slow: the deviation from
the stationary state decays non-exponentially in time, as a power law with an exponent that depends continuously on the system parameters, namely
$r$ and $D$~\cite{BM2024_1}. This result for the uniform kernel in the presence of a confining potential was recently generalized to the case
of wider range of kernels~\cite{BEM2025}.  
On the mathematical side, a central limit theorem and an anomalous 
large deviation principle have also been established for a class of memory walks of this type for a broad range of memory 
kernels~\cite{MU2019,BM2023}. The rigorous proofs of these results are based on a connection between the resetting process to 
previous sites and the growth of weighted random recursive trees.
In a more applied context, random walks with long range memory have become increasingly useful in ecology for the description and 
analysis of animal mobility. There is mounting evidence that animals do not follow purely Markov processes but use their memory and 
tend to revisit preferred places during ranging \cite{BDF2008,F2008,VM2009,F2013,MFM2014,F2023}. The model above was able to 
describe quantitatively the movement patterns of Capuchin monkeys in the wild \cite{BS2014}, as well as of individual elks released 
in an unknown environment \cite{FC2021}. Another feature shared by many animal trajectories is spatial localization, by opposition 
to unbounded diffusion, as home ranges are often clearly identifiable.
The range-resident behaviour of animals can be modelled by incorporating a central place toward which 
they are attracted (such as a den), which produces an effective piece-wise linear 
potential and therefore a stationary 'equilibrium' state \cite{ML2006}. Recent studies have 
considered harmonic potentials and Ornstein–Uhlenbeck (OU) processes as the starting point of models of 
animal motion of increasing complexity \cite{F2017,MG2020,F2022}.

\begin{figure}[t]
\centering
\includegraphics[width=.5\textwidth,angle=-90]{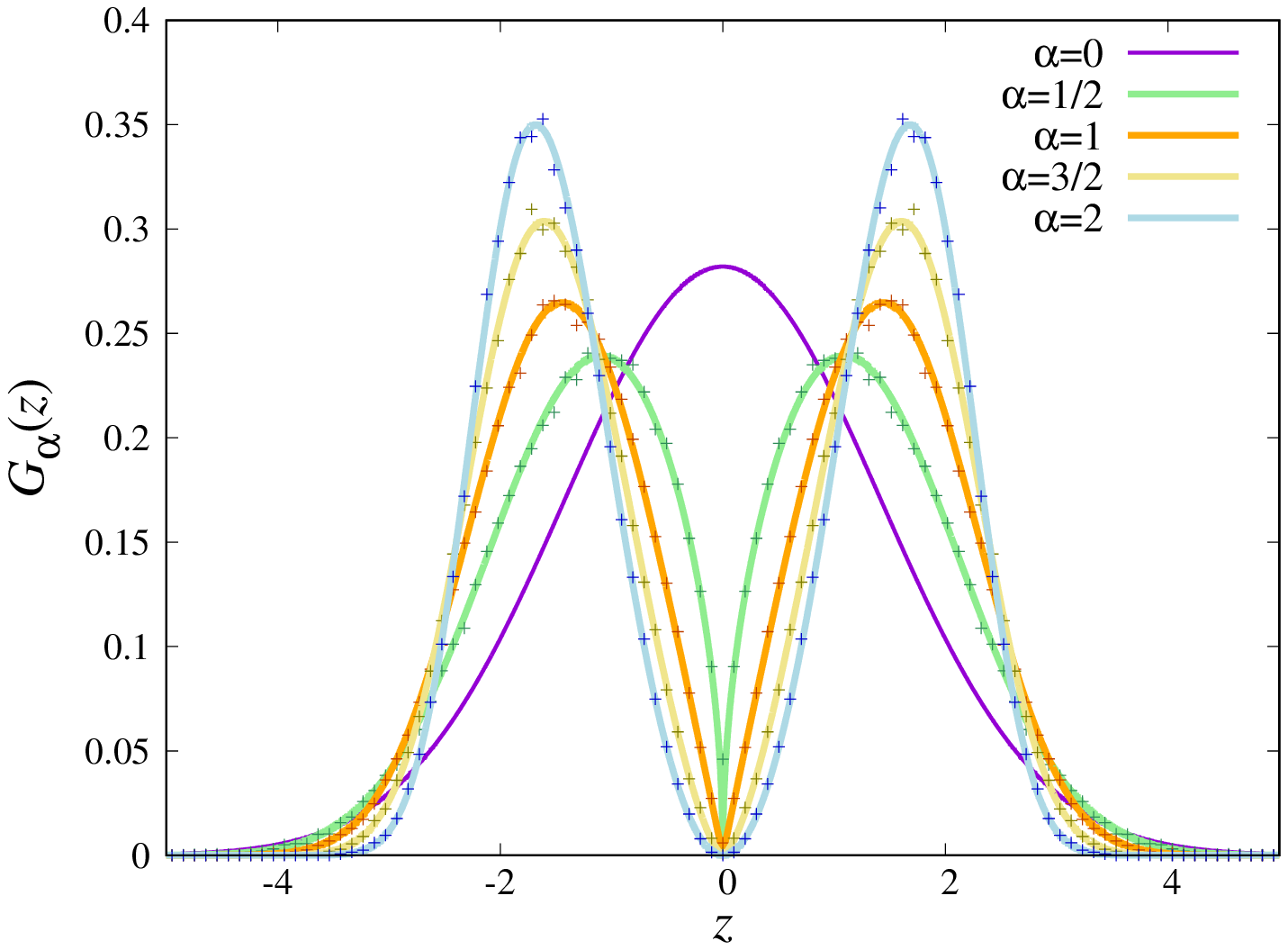}
    \caption{Scaling functions of the sluggish random walker in one dimension, given by Eq. (\ref{scaling.1}), for various values of $\alpha$ (solid lines). Results obtained from simulations (see  \ref{app:simul} for details) are shown with symbols for comparison.  The walker follows a Brownian motion where the diffusion coefficient decreases with the distance from the origin as $D(x)=|x|^{-\alpha}$.}
    \label{fig:scal}
\end{figure}

Another simple random walk lattice model leading to subdiffusive growth at late times has been introduced recently, known
as the `sluggish random walk' (SRW)~\cite{ZAEM2023}. In the SRW model, e.g., in one dimension, the hopping rate out of a site located
at $x$ decreases as $|x|^{-\alpha}$ as $|x|$ increases (with $\alpha>0$), making the walker sluggish the further it goes from the
origin. In the continuum limit of 
this model,
the position of the walker evolves via the Langevin equation
\begin{equation}
\frac{dx}{dt}= \sqrt{2\, D(x)}\, \eta(t)\, ,
\label{lange.1}
\end{equation}
where $\eta(t)$ is a zero mean Gaussian white noise with delta correlations: $\langle \eta(t)\eta(t')\rangle=\delta(t-t')$
and $D(x) \sim |x|^{-\alpha}$ is a space dependent diffusion constant that falls off algebraically
with the distance from the origin. The Langevin equation (\ref{lange.1}) is interpreted
in the  It\^{o} sense, since in the associated lattice model the hopping rate depends only on the departure site, and not on the destination site. The associated Fokker-Planck equation for the position distribution reads~\cite{ZAEM2023}
\begin{equation}
\frac{\partial P(x,t)}{\partial t}= \frac{\partial^2}{\partial x^2}\left[D(x)\, P(x,t)\right]\, ,
\label{fp.1}
\end{equation}
an equation also studied with $D(x)=|x|^{-\alpha}$ in Ref. \cite{FL2003}.
Note that $D(x)$ appears inside the Laplacian in Eq. (\ref{fp.1}) due to the It\^{o} rule.
The Fokker-Planck equations in different interpretations and their solutions obtained with $D(x)=|x|^{-\alpha}$ have also been discussed in the literature ~\cite{CCM2013,LB2019,SCT2023}.
The choice $D(x) \sim |x|^{-\alpha}$ for $|x|\gg 1$ creates an effective drift away from the origin, 
but the diffusion rate also falls off as a power law with
distance. The solution of the Fokker-Planck equation (\ref{fp.1}) admits a scaling form at late times~\cite{FL2003,ZAEM2023},
\begin{equation}
P(x,t) \approx \frac{1}{t^{\nu}}\, G_{\alpha}\left( \frac{x}{t^{\nu}}\right)\, \quad {\rm with}\quad \nu=\frac{1}{\alpha+2}\, ,
\label{position_dist.1}
\end{equation}
where $G_{\alpha}(z)$ is a symmetric but non-Gaussian scaling function (for any $\alpha>0$). It is actually non-monotonous with respect to $|z|$ and given by~\cite{FL2003,ZAEM2023}
\begin{equation}
G_{\alpha}(z)= A_{\alpha}\, |z|^ {\alpha}\, \exp\left(-\nu^2\, |z|^{1/\nu}\right)\, .
\label{scaling.1}
\end{equation}
The normalization constant $A_{\alpha}$ can be explicitly computed as
\begin{equation}
A_{\alpha}= \frac{\nu^{1-2\nu}}{2\, \Gamma(1-\nu)}\, \quad {\rm with}\quad \nu=\frac{1}{\alpha+2}\, .
\label{norm.1}
\end{equation}
The scaling form (\ref{position_dist.1}) of $P(x,t)$ shows that the typical position of
the walker spreads subdiffusively as $x\sim t^{1/(\alpha+2)}$, recovering the diffusive spread only for $\alpha=0$. The scaling function $G_{\alpha}(z)$ is shown in Fig. \ref{fig:scal}.
Various other observables such as the survival probability up to time $t$ and the distribution of the maximal
displacement up to time $t$ can be computed explicitly for all $\alpha\ge 0$~\cite{ZAEM2023}. Recently, distributions of other functionals
of this process such as the occupation time, the time at which the maximum occurs and the last-passage time before $t$
have been computed explicitly, leading to non-trivial generalization (for $\alpha>0)$ of the arcsine laws valid 
for $\alpha=0$~\cite{DM2025}.

Diffusion of a particle with a space-dependent diffusion coefficient $D(x)=|x|^{-\alpha}$ and subject to 
stochastic resetting to a fixed position (with rate $r$) has been studied in the anti-It\^o interpretation 
\cite{LLGER2022}. Due to resetting, the position distribution tends asymptotically to a non-equilibrium 
steady state, which can be obtained exactly, as well as the mean square displacement at all times. The 
mean first-passage time of the particle at the origin can also be calculated \cite{LLGER2022}. Some exact 
results on first-passage observables can also be obtained for an arbitrary profile $D(x)$ in the 
Stratonovitch sense, while approximate expressions at large and small $r$ are available in the other 
interpretations \cite{MA2024}.
The first-passage properties of the related Feller process (where $\alpha=-1$ or $D(x)=|x|$) subject to stochastic resetting have been derived in \cite{R2022}.
The sluggish random walker model with stochastic resetting
to the initial position has also been studied recently in the context of an adaptive and efficient search 
strategy called `proxitaxis'~\cite{DKMS2025}.


In summary, both the memory dependent resetting with rate $r$ and the sluggishness in the diffusion 
via $D(x)\sim  |x|^{-\alpha}$ lead to a slowdown of the late time dynamics of the random walker.
In the former case, the typical distance of the walker grows as $\sim \sqrt{\ln (rt)}$ at late times even though the position distribution
remains Gaussian. In the latter case, the typical distance grows as a subdiffusive power law $\sim t^{1/(\alpha+2)}$, but
the position distribution becomes highly non-Gaussian for any $\alpha>0$. It is then natural to ask what happens to
a sluggish random walker with $D(x)\sim |x|^{-\alpha}$, if in addition it undergoes memory dependent resetting with rate $r$.
How does the typical position grow with time and what is the position distribution at late times? In this paper,
we provide an exact solution for the position distribution in this model with both sluggishness and memory dependent resetting.
We show that these two moves lead to an ultra slowdown in the dynamics and the typical distance of the walker
from the origin grows sub-logarithmically, $x\sim \left[\ln (rt)\right]^{1/(\alpha+2)}$ at late times. However, the
scaling function describing the position distribution remains the same as in Eq. (\ref{scaling.1}), i.e., unchanged from
the scaling function $G_{\alpha}(z)$  of the sluggish walker without resetting.

The rest of the paper is organized as follows. In Section ~\ref{model}, we define the
model precisely and derive the exact position distribution at all times in one dimension.
In Section ~\ref{moments}, we derive the exact generalized moments of the position distribution
and verify the analytical predictions numerically. 
Section \ref{sec:msd} presents the asymptotic behaviour of the variance, from which a finite time approximation is derived.
Section~\ref{higher_d} provides
the derivation of the exact position distribution in arbitrary dimensions. We conclude with
a summary and outlook in Section~\ref{summary}. Details of the numerical simulations are
provided in ~\ref{numerical}.

\section{The model and its exact solution in one dimension}
\label{model}

We start with a sluggish random walker on a line evolving via the Langevin equation (\ref{lange.1}) with
a space dependent diffusion rate $D(x)$ (symmetric in $x$), starting at the origin at $t=0$.
For the moment we consider a general $D(x)$, but eventually we will focus on 
$D(x)= |x|^{-\alpha}$. With this choice, the model remains well behaved 
in the continuum even in the vicinity of $x\to 0$.
We now incorporate the memory driven resetting moves to this sluggish dynamics.   
At time $t$, with rate $r$ the particle chooses any previous time in the past $0\le t'< t$ with
uniform probability density $1/t$ and resets to the position that it occupied at that previous instant $t'$. Let $P_r(x,t)$
denote the position distribution at time $t$. To take into account the memory effect, we now need to define the
two-point function $P_r(x,t\,; x', t')$, which is the joint probability density for the particle to be at $x'$ at $t'$
and at $x$ at $t$, with $t'\le t$. Clearly, if we integrate over one of the positions, we recover the
marginal one-point probability density
\begin{equation}
\int_{-\infty}^{\infty} P_r(x,t\, ; x', t')\, dx'= P_r(x,t)\,  \ {\rm and}\ \int_{-\infty}^{\infty}P_r(x,t\, ; x', t')\, dx= P_r(x',t')\, .
\label{marg.1}
\end{equation}
Subsequently, the Fokker-Planck equation for the evolution of the one-point
position distribution $P_r(x,t)$ can be written as
\begin{equation}
\partial_t P_r(x,t)= \partial_x^2 \left[D(x)\, P_r(x,t)\right] - r\, P_r(x,t) +
\frac{r}{t}\, \int_0^t dt' \int_{-\infty}^{\infty} dx'\, P_r(x',t; x, t')\, .
\label{fpr.1}
\end{equation}
The first term on the right hand side (rhs) describes the space-dependent diffusion via the It\^{o} rule.
The second term
describes the loss of probability density from position $x$ at time $t$ due to resetting to other positions.
The last term on the rhs describes the gain in the probability density at $x$ at time $t$ due to resetting
from other positions labelled by $x'$. If the particle has to reset to $x$ from $x'$ at time $t$, it must have been
at $x$ at a previous time $t'\le t$ and the probability density of this event is simply the
two-point function $ P_r(x',t; x, t')$. Finally, we need to integrate over all times $t'$ where $x$ may have been visited, and all positions $x'$ from
which the particle may arrive at $x$ at time $t$ via resetting. The reason for the solvability for the one-point
position distribution in this model can then traced back to the second relation in Eq. (\ref{marg.1}), which
allows us to write a closed equation for the one-point function (without involving multiple-point functions). 
We then obtain
\begin{equation}
\partial_t P_r(x,t)=\partial_x^2 \left[D(x)\, P_r(x,t)\right] - r\, P_r(x,t) +
\frac{r}{t}\, \int_0^t dt' \,  P_r(x, t')\, .
\label{fpr.2}
\end{equation}
By integrating Eq. (\ref{fpr.2}) over $x$ and if no absorbing boundaries are present, it is easy to check that 
$\int_{-\infty}^{\infty} P_r(x,t)\, dx$ is independent of time, i.e., the total probability is 
conserved and equals unity since the initial condition, e.g., $P_r(x,0)= \delta(x-x_0)$ normalizes the
probability to unity. One also needs to impose the vanishing boundary conditions as $x\to \pm \infty$, i.e.,
$P_r(x\to \pm \infty, t)=0$ for all $t$. 
Evidently, for $r=0$, Eq. (\ref{fpr.2}) reduces to the 
Fokker-Planck equation (\ref{fp.1}). Note that $P_r(x,t)$ of course depends on the choice of $D(x)$, but we
suppress the explicit dependence of $P_r(x,t)$ on $D(x)$ for the moment to keep notational simplicity, and
we will restore it later at an appropriate place.

To solve the Fokker-Planck equation (\ref{fpr.2}), we use the
method of separation of variables with the ansatz
\begin{equation}
P_r(x,t) = \phi(x)\, f_r(t)\, .
\label{sepr.1}
\end{equation}
Substituting (\ref{sepr.1}) in Eq. (\ref{fpr.2}), dividing both sides by $\phi(x) f_r(t)$ and assembling
the only-time dependent and only-space dependent parts separately gives
\begin{equation}
\frac{{\dot f_r}(t)}{f_r(t)} + r - \frac{r}{t\, f_r(t)}\, \int_0^t f_r(t')\, dt' = 
\frac{1}{\phi(x)}\, \frac{d^2}{dx^2} \left[ D(x)\,\phi(x)\right]= -\lambda\, ,
\label{sepr.2}
\end{equation}
where $\lambda\ge 0$ is a constant independent of $x$ and $t$. Note that the time-dependent part
$f_r(t)$ is completely independent of $D(x)$, while the
space-dependent function $\phi(x)$ is independent of $r$. We now solve the time and space dependent parts
separately.

\subsection{Time-dependent part}
In fact, the time-dependent part $f_r(t)$, which is independent of $D(x)$, was already solved in Ref.~\cite{BM2024_1}, which
we recall briefly here. For a given $\lambda$, we get from Eq. (\ref{sepr.2})
a second order ordinary differential equation for $f_r(t)$
\begin{equation}
{\dot f_r}(t) + (r+\lambda)\, f_r(t)- \frac{r}{t}\, \int_0^t f_r(t')\, dt'=0\, .
\label{ft_diff.1}
\end{equation}
Multiplying this equation by $t$ and taking the time derivative, one obtains
\begin{equation}
t{\ddot f_r}(t) + [1+(r+\lambda)t]\, \dot{f_r}(t)+ \lambda f_r(t)=0\, .
\label{cumul_Ft.1}
\end{equation}
Through the change of variable $z=-(\lambda+r)t$ with $g(z)\equiv f(t)$, Eq. (\ref{cumul_Ft.1}) can be recast as a confluent hypergeometric equation,
\begin{equation}
z\frac{d^2 g(z)}{dz^2}+(b-z)\frac{dg(z)}{dz}-ag(z)=0,
\label{Ft_diff.2}
\end{equation}
with $b=1$ and $a=\lambda/(\lambda+r)$. The differential equation (\ref{Ft_diff.2}) has two 
linearly independent solutions denoted by $M(a,b,z)$ and $U(a,b,z)$ \cite{AS_book}.
The function $M(a,b,z)$ is known as Kummer's function and has a simple power series expansion
\begin{equation}
M(a,b,z)= 1+ \frac{a}{b}\, z + \frac{a(a+1)}{b(b+1)}\, \frac{z^2}{2!}+ \ldots\, .
\label{Kummer_def}
\end{equation}
Its linearly independent cousin $U(a,b,z)$ is called the confluent hypergeometric function and
has a more complicated expression~\cite{AS_book}.
Hence, the most general solution of Eq. (\ref{cumul_Ft.1}) can be expressed as
\begin{equation}
f_r(t) = c_1\,  M \left( \frac{\lambda}{r+\lambda}, 1, - (r+\lambda) t\right)
+ c_2\,\, U \left( \frac{\lambda}{r+\lambda}, 1, - (r+\lambda) t\right)\, ,
\label{gen_sol.1}
\end{equation}
where $c_1$ and $c_2$ are two arbitrary constants. The solution must be finite at all $t$, in particular at $t=0$. The small argument asymptotics of the two solutions are
given by~\cite{AS_book}
\begin{eqnarray}
M(a,1,z) &\to & 1\, \quad\quad\quad\quad\quad\ \,\, {\rm as}\ z\to 0 \label{Mz0.1} \\
U(a,1,z) &\to & -\frac{1}{\Gamma(a)}\, \ln z \, \quad {\rm as}\ z\to 0 \, . 
\label{Uz0.1} 
\end{eqnarray}
Consequently, the logarithmic divergence as $t\to 0$ in Eq. (\ref{gen_sol.1}) is avoided by setting $c_2=0$.
Then the solution simply reads
\begin{equation}
f_r(t) = M \left( \frac{\lambda}{r+\lambda}, 1, - (r+\lambda) t\right)\, .
\label{ft_sol.1}
\end{equation} 
where we have set $c_1=1$ by absorbing this constant into the amplitude of
the space-dependent eigenmode $\phi(x)$.

\subsection{Space-dependent part} 
The space dependent part $\phi(x)$ in Eq. (\ref{sepr.2}) evidently depends on $\lambda$. Hence
we denote it by $\phi_{\lambda}(x)$ and it satisfies the eigenvalue equation
\begin{equation}
\frac{d^2}{dx^2} \left[D(x)\, \phi_\lambda(x)\right] + \lambda\, \phi_{\lambda}(x)=0\, .
\label{space_part.1}
\end{equation}
Note that since $D(x)$ is symmetric in $x$, so is $\phi_{\lambda}(x)$. 
To proceed further, we need to
make a choice for $D(x)$. Following the sluggish random walker model 
studied in Ref.~\cite{ZAEM2023},
we choose $D(x)= |x|^{-\alpha}$. We next define 
\begin{equation}
h_{\lambda}(x)= \frac{\phi_{\lambda}(x)}{|x|^{\alpha}}\, ,
\label{hx_def}
\end{equation}
which satisfies the ordinary second order differential equation 
\begin{equation}
\frac{d^2 h_{\lambda}(x)}{dx^2}+ \lambda\, |x|^{\alpha}\, h_{\lambda}(x)=0\, .
\label{hx_diff.1}
\end{equation}
Since $h_\lambda(x)$ is symmetric around $x=0$, we will henceforth consider only $x\ge 0$.
This differential equation (\ref{hx_diff.1}) can fortunately be solved exactly for any 
$\alpha\ge 0$. The general solution for $x\ge 0$ reads
\begin{equation}
h_{\lambda}(x)= B_1 (\lambda)\, \sqrt{x}\, J_{-\nu}\left( 2\, \nu\, \sqrt{\lambda}\, x^{1/(2\nu)}\right)+
B_2(\lambda)\, \sqrt{x}\ J_{\nu}\left(2\, \nu\, \sqrt{\lambda}\, x^{1/(2\nu)}\right)\, , 
\label{h_sol.1}
\end{equation}
where
\begin{equation}
\nu= \frac{1}{\alpha+2}\, .
\end{equation}
Here $J_{\pm \nu}(z)$ is the standard Bessel function of index $\pm \nu$ and argument $z$. The two unknown 
constants (i.e., independent of $x$) $B_1(\lambda)$
and $B_2(\lambda)$ need to be fixed from the boundary conditions. Let us first consider the 
boundary $x=0$. It is clear from Eq. (\ref{hx_def}) that $h_{\lambda}(x)$ is symmetric in $x$
and its second derivative must exist as $x\to 0$.
Let us now take the $x\to 0$ limit of
the solution (\ref{h_sol.1}) using the following asymptotic behaviour of the Bessel function~\cite{AS_book}
\begin{equation}
J_\nu(z)\to \frac{2^{-\nu}}{\Gamma(\nu+1)}\, z^{\nu}\quad{\rm as}\ z\to 0\, .
\label{Jz_asymp.1}
\end{equation} 
This gives
\begin{equation}
h_{\lambda}(x) \to \left[\frac{\nu^{-\nu}\, \lambda^{-1/(2\nu)}}{\Gamma(1-\nu)}\right]\, B_1(\lambda)
+ \left[\frac{\nu^{\nu}\, \lambda^{1/(2\nu)}}{\Gamma(1+\nu)}\right]\, B_2(\lambda)\, x \, \quad {\rm as}\ x\to 0 \, .
\label{h_x0.1}
\end{equation} 
Since $h_{\lambda}(x)$ is symmetric in $x$, it can not have a linear slope as $x\to 0$, since that would lead
to a cusp at $x=0$ and the second derivative $h_{\lambda}''(0)$ will not exist. To rule out the linear slope at
$x=0$, we therefore need to set $B_2(\lambda)=0$, leading to
\begin{equation}
h_{\lambda}(x)= B_1(\lambda)\, \sqrt{x}\, J_{-\nu}\left( 2\, \nu\, \sqrt{\lambda}\, x^{1/(2\nu)}\right)\, .
\label{h_sol.2}
\end{equation}
Consequently, the space dependent part of the
eigensolution $\phi_{\lambda}(x)$, using Eq. (\ref{hx_def}), is given by
\begin{equation} 
\phi_{\lambda}(x)= B_1(\lambda)\, x^{\alpha+1/2}\, J_{-\nu}\left( 2\, \nu\, \sqrt{\lambda}\, x^{1/(2\nu)}\right)\, ,
\label{phi_sol.1}
\end{equation}
where we recall that $\nu=1/(\alpha+2)$. 
Since $J_\nu(z)\sim  \cos(z-\pi\nu/2-\pi/4)/\sqrt{z}$ as $z\to \infty$, it follows
that $\phi_{\lambda}(x)$ in Eq. (\ref{phi_sol.1}) diverges as $x^{3\alpha/4}$ at large $x$.
This, however, is not a problem for the full solution $P_r(x,t)$ when all
the modes are summed over (see the next subsection). This is due to non-trivial cancellations at large argument that
occur due to the oscillatory nature of the Bessel function.

\subsection{The full solution}

For any fixed $\lambda$, we have now determined both the time dependent part $f_r(t)$ and the space dependent part $\phi_{\lambda}(x)$
of the eigensolution characterized by the eigenvalue $\lambda\ge 0$. Their product, for any fixed $\lambda\ge 0$, is a solution of the
original Fokker-Planck equation. In fact the most general solution of the Fokker-Planck equation (\ref{fpr.2}) can be
expressed as a linear combination over all possible values of $\lambda\ge 0$. Since the position distribution now
depends on the two parameters $(r,\alpha)$ and the variables $x$ and $t$, we will denote, henceforth, the full solution
by $P_{r,\alpha}(x,t)$ to make the dependence on the two parameters explicit.

Hence,
using the time-dependent part $f_r(t)$ in Eq. (\ref{ft_sol.1})
and the space-dependent part $\phi_{\lambda}(x)$ in Eq. (\ref{phi_sol.1}),  we can express, for any $t$ and any $x\ge 0$, 
the full solution $P_{r,\alpha}(x,t)$ in the form of
the following exact spectral decomposition
\begin{equation}
\fl
P_{r,\alpha}(x,t)= x^{\alpha+1/2}\, \int_0^{\infty} d\lambda\, B_1(\lambda)\, J_{-\nu}\left( 2\, \nu\, \sqrt{\lambda}\, x^{1/(2\nu)}\right)\,  
M \left( \frac{\lambda}{r+\lambda}, 1, - (r+\lambda) t\right)\, , 
\label{Prxt_sol.1}
\end{equation}
where
\begin{equation}
\nu=\frac{1}{\alpha+2}\, .
\end{equation}
The only unknown quantity is the spectral density $B_1(\lambda)$ which is determined from the initial condition as follows. 
We assume that initially the particle is located at $x=0$.
Since the solution $P_{r,\alpha}(x,t)$, with this initial condition is symmetric in $x$, we must have
$\int_0^{\infty} P_{r,\alpha}(x,t) dx=1/2$. In particular, at $t=0$ and focusing only on $x\ge 0$, we have
\begin{equation}
P_{r,\alpha}(x,0)= \frac{1}{2}\, \delta(x)\, .
\label{init_cond.1}
\end{equation}
We set $t=0$ in Eq. (\ref{Prxt_sol.1}) and use $M(a,b,z=0)=1$. Comparing it to 
(\ref{init_cond.1}) then leads to
\begin{equation}
x^{\alpha+1/2}\, \int_0^{\infty} d\lambda\, B_1(\lambda)\, J_{-\nu}\left( 2\, \nu\, \sqrt{\lambda}\, x^{1/(2\nu)}\right)= 
\frac{1}{2}\, \delta(x)\, .
\label{sp_dens.1}
\end{equation}

To determine $B_1(\lambda)$ from Eq. (\ref{sp_dens.1}) we proceed as follows. We first multiply both sides of (\ref{sp_dens.1})
by the factor $\sqrt{x}\, J_{-\nu}\left( 2\,\nu\, \sqrt{\lambda'}\, x^{1/(2\nu)}\right)$ with $\lambda'>0$ and integrate over $x\in [0,\infty]$.
On the rhs this integral gives
\begin{equation}
{\rm rhs} = \frac{1}{2}\, \lim_{x\to 0} \sqrt{x}\, J_{-\nu}\left( 2\,\nu\, \sqrt{\lambda'}\, x^{1/(2\nu)}\right)\, .
\label{rhs_lim.1}
\end{equation}
Using the asymptotic behaviour of the Bessel function in Eq. (\ref{Jz_asymp.1}) we then get
\begin{eqnarray}
&&\int_0^{\infty} d\lambda\, B_1(\lambda)\, \int_0^{\infty} dx\, x^{\alpha+1}\, J_{-\nu}\left( 2\,\nu\, \sqrt{\lambda'}\, x^{1/(2\nu)}\right)\,
J_{-\nu}\left( 2\,\nu\, \sqrt{\lambda}\, x^{1/(2\nu)}\right)\nonumber\\
&&= \frac{1}{2}\, \frac{\nu^{-\nu}}{\Gamma(1-\nu)}\, \left(\lambda'\right)^{-\nu/2}\, , 
\label{sp_dens.2}
\end{eqnarray}
where we recall that $\nu= 1/(\alpha+2)$. Making the substitution $2 \nu x^{1/(2\nu)}= y$ and simplifying the integral, we get
\begin{equation}
\int_0^{\infty} d\lambda\, B_1(\lambda)\, \int_0^{\infty} dy\, y\, J_{-\nu}\left(\sqrt{\lambda'}\, y\right)\, 
J_{-\nu}\left(\sqrt{\lambda}\, y\right)= \frac{\nu^{1-\nu}}{\Gamma(1-\nu)}\, \left(\lambda'\right)^{-\nu/2}\, .
\label{sp_dens.3}
\end{equation}
To extract $B_1(\lambda)$, we next use the following identity~\cite{Jackson_book}
\begin{equation}
\int_0^{\infty} dy\, y\, J_{m}(k y)\, J_m(k'\, y) =\frac{1}{k}\, \delta(k-k')\, ,
\label{identity_jackson.1}
\end{equation}
valid for any $m$. 
Setting $k= \sqrt{\lambda}$ and $m=-\nu$ gives
\begin{equation}
\int_0^{\infty} dy\, y\, J_{-\nu}\left(\sqrt{\lambda}\, y\right)\, J_{-\nu}\left(\sqrt{\lambda'}\, y\right) =
2\,\delta\left(\lambda-\lambda'\right)\, .
\label{identity_jackson.2}
\end{equation}
Using this identity in Eq. (\ref{sp_dens.3}) fixes $B_1(\lambda)$ as
\begin{equation}
B_1(\lambda)= \frac{\nu^{1-\nu}}{2\, \Gamma(1-\nu)}\, \lambda^{-\nu/2}\, \quad {\rm with}\quad \nu=\frac{1}{\alpha+2}\, .
\label{b1l_sol.1}
\end{equation}
Plugging this expression for $B_1(\lambda)$ in Eq. (\ref{Prxt_sol.1}) gives the 
full exact position distribution, valid for all
$x\ge 0$ (and equivalently for $x<0$ by symmetry) and all time $t$
\begin{eqnarray}
\fl
P_{r,\alpha}(x,t)=&& \label{Prxt_sol.2}\\
&&\fl\frac{\nu^{1-\nu}}{2\, \Gamma(1-\nu)}\, x^{\alpha+1/2}\, \int_0^{\infty} d \lambda\, \lambda^{-\nu/2}\,
J_{-\nu}\left( 2\nu\sqrt{\lambda}\, x^{1/(2\nu)}\right)\, 
M \left( \frac{\lambda}{r+\lambda}, 1, - (r+\lambda) t\right)\, . \nonumber
\end{eqnarray}

As a useful check, let us first demonstrate that our general exact solution (\ref{Prxt_sol.2}) valid for arbitrary $r\ge 0$ does 
reproduce the known solution for $r=0$~\cite{ZAEM2023}. To proceed, we first note that for $r=0$, the time-dependent part
of the solution in Eq. (\ref{Prxt_sol.2}) greatly simplifies since
\begin{equation}
\lim_{r\to 0} M \left( \frac{\lambda}{r+\lambda}, 1, - (r+\lambda) t\right)= M\left(1,1, -\lambda t\right)= e^{-\lambda\, t}\, .
\label{time_r0.1}
\end{equation}
This last identity follows from the definition of $M(a,b,z)$ in Eq. (\ref{Kummer_def}). Hence, in the limit $r\to 0$, our result
(\ref{Prxt_sol.2}) reduces to
\begin{equation}
P_{0,\alpha}(x,t)= \frac{\nu^{1-\nu}}{2\, \Gamma(1-\nu)}\, x^{\alpha+1/2}\, \int_0^{\infty} d \lambda\, \lambda^{-\nu/2}\,
J_{-\nu}\left( 2\,\nu\, \sqrt{\lambda}\, x^{1/(2\nu)}\right)\, e^{-\lambda\, t}\, .
\label{P0xt_sol.1}
\end{equation}
Amazingly, the integral over $\lambda$ can be done explicitly using the identity (Mathematica can perform 
this integral explicitly) 
\begin{equation}
\fl
\int_0^{\infty} d\lambda\, \lambda^{-\nu/2}\, J_{-\nu} \left(2\,\nu\, \sqrt{\lambda}\, x^{1/(2\nu)}\right)\, e^{-\lambda\, t}
= \nu^{-\nu}\, x^{-1/2}\, t^{\nu-1}\, \exp\left(-\frac{\nu^2}{t}\, x^{1/\nu}\right)\, ,
\label{iden.2}
\end{equation}
valid for any $\nu>0$ and $x\ge 0$. Using this identity, we get from Eq. (\ref{P0xt_sol.1})
\begin{equation}
P_{0,\alpha}(x,t)= \frac{\nu^{1-2\nu}}{2\, \Gamma(1-\nu)}\, t^{\nu-1}\, x^{\alpha}\, e^{-\frac{\nu^2}{t}\, x^{1/\nu}}\, \quad 
{\rm where}\quad \nu=\frac{1}{\alpha+2}\, .
\label{P0xt_sol.2}
\end{equation}
The solution (\ref{P0xt_sol.2}) can be written in the scaling form 
\begin{equation}
P_{0,\alpha}(x,t) = \frac{1}{t^{\nu}}\, G_\alpha\left( \frac{x}{ t^{\nu}}\right)\, ,
\label{P0xt_scaling.1}
\end{equation}
where the scaling function $G_{\alpha}(z)$ has precisely the same expression as given in Eq. (\ref{scaling.1}).
This then reproduces the known result in the limit $r=0$.
Indeed, this result for $r=0$ was derived in Ref.~\cite{ZAEM2023} by directly substituting the scaling ansatz (\ref{P0xt_scaling.1}) in the Fokker-Planck
equation (\ref{fp.1}) (with $D(x)= |x|^{-\alpha}$) and solving the resulting ordinary differential equation for $G_\alpha(z)$.
Here, our method using spectral decomposition thus provides an alternative derivation of this result for $r=0$.

Another important check corresponds to choosing $\alpha=0$, i.e., $\nu=1/2$, where the exact result for
$P_{r,0}(x,t)$ is known~\cite{BEM2017}. Setting $\nu=1/2$ in our general result in Eq. (\ref{Prxt_sol.2}), we get
\begin{equation}
\fl
P_{r,0}(x,t)= \frac{1}{2\sqrt{2\pi}}\, \sqrt{x}\, \int_0^{\infty} d\lambda\, \lambda^{-1/4}\, 
J_{-1/2}\left(\sqrt{\lambda}\, x\right)\,  M\left( \frac{\lambda}{r+\lambda}, 1, - (r+\lambda) t\right)\, .
\label{a0_sol.1}
\end{equation}
Setting $\lambda=k^2$ and using $ J_{-1/2}(k x)= \sqrt{\frac{2}{\pi k x}}\, \cos(kx)$ gives
\begin{equation}
P_{r,0}(x,t)= \frac{1}{\pi}\, \int_0^{\infty} dk\, \cos(kx)\, M\left( \frac{k^2}{r+k^2}, 1, - (r+k^2) t\right)\, ,
\label{a0_sol.2}
\end{equation}
which coincides with the solution found in Ref.~\cite{BEM2017} by solving directly the Fokker-Planck equation
in the Fourier space.

Thus, our general result for $P_{r,\alpha}(x,t)$ in Eq. (\ref{Prxt_sol.2}), 
valid for arbitrary $r$ and $\alpha\ge 0$,
contains both the known limiting results when $r\to 0$ and $\alpha\to 0$. 
Let us also remark that although here we assumed $\alpha\ge 0$ to model
the `sluggishness' of the walker (so that the diffusion coefficient decreases with
increasing $|x|$), the result for the position distribution in Eq. (\ref{Prxt_sol.2})
is actually valid for any $\alpha>-1$. For $\alpha<-1$, the solution in Eq. (\ref{Prxt_sol.2})
is not normalizable.

\subsection{Asymptotic behaviour at late times for general $r$ and $\alpha$}

Let us now derive the asymptotic late time behaviour of the position distribution $P_{r,\alpha}(x,t)$ in Eq. (\ref{Prxt_sol.2}).
At late times, the argument $-(r+\lambda) t$ of $M(a,b, -(r+\lambda)\, t)$ in Eq. (\ref{Prxt_sol.2}) becomes a large
negative number. We can then use the following asymptotic behaviour~\cite{AS_book}
\begin{equation}
M(a,b,-z) \approx \frac{\Gamma(b)}{\Gamma(b-a)}\, z^{-a}\,  \quad {\rm as}\quad z\to \infty\, .
\label{M_asymp.1}
\end{equation}
Consequently, 
\begin{eqnarray}
M\left(\frac{\lambda}{r+\lambda}, 1, - (r+\lambda) t\right)&\approx& \frac{1}{\Gamma\left(\frac{r}{r+\lambda}\right)}\, 
\left[(r+\lambda) t\right]^{- \lambda/(r+\lambda)}\nonumber\\
&=&\frac{1}{\Gamma\left(\frac{r}{r+\lambda}\right)}\, e^{- \frac{\lambda}{r+\lambda}\,
\ln\left( (r+\lambda)t\right)}\, .
\label{M_asymp.2}
\end{eqnarray}
We substitute this large $t$ behavior in Eq. (\ref{Prxt_sol.2}),
rescale $\lambda= r\, \tilde{\lambda}/\ln(r t)$ and at the same time rescale also
$x= \left(\frac{1}{r}\, \ln(rt)\right)^{\nu}\, \tilde{x}$ in the integral.
The resulting rescaled integral over $\tilde{\lambda}$ (with $\tilde{x}$ fixed) can be performed exactly using
the identity (\ref{iden.2}). Taking into account all factors carefully, it is then easy to see that $P_r(x,t)$, for large $t$
and large $x$, but with the ratio $\tilde{x}= x\, r^{\nu}/\left(\ln (rt)\right)^{\nu}$ fixed, approaches the following scaling form
\begin{equation}
P_{r,\alpha}(x,t) \approx \left(\frac{r}{\ln (rt)}\right)^{\nu}\, G_{\alpha}\left( \frac{r^\nu\, x}{\left[\ln (rt)\right]^{\nu}}\right)\, \quad
{\rm with} \quad \nu=\frac{1}{\alpha+2}\, ,
\label{Prxt_asymp.1}
\end{equation}
where the scaling function $G_{\alpha}(z)$ is precisely the one given in Eq. (\ref{scaling.1}). Thus the effect of a non-zero resetting
rate $r$ is to slow down the dynamics drastically with $x\sim \left(\ln (rt)\right)^{1/(\alpha+2)}$ at late times instead of $t^{1/(\alpha+2)}$, but
remarkably the associated scaling function $G_{\alpha}(z)$ retains the same form as $r=0$.

\section{Generalized moments of the position distribution}
\label{moments}

The spectral representation of the position distribution in Eq. (\ref{Prxt_sol.2}) is valid at all $t$ and for all $r$ and 
$\alpha\ge 0$. It helped us to derive explicitly the late time scaling behaviour of $P_{r,\alpha}(x,t)$ in Eq. (\ref{Prxt_asymp.1}).
However, plotting $P_r(x,t)$ at finite $t$ using the spectral decomposition (\ref{Prxt_sol.2}) is not easy, since the integral over
$\lambda$ is difficult to do explicitly. Hence, it would be nice to find an observable that can be explicitly computed at all
$t$ and compared to numerical simulations. A natural candidate for such an observable is the quantity
\begin{equation}
\mu_{r,\alpha}(m,t)= \int_{-\infty}^{\infty} P_{r,\alpha}(x,t)\, |x|^{m}\, dx= 2 \int_0^{\infty} P_{r,\alpha}(x,t)\, |x|^{m}\, dx\, ,
\label{mom_ra.1}
\end{equation}
where we used the symmetry $P_{r,\alpha}(-x,t)= P_{r,\alpha}(x,t)$ of the distribution. For $m$ an even integer, 
$\mu_{r,\alpha}(m,t)$ is indeed the $m$-th integer moment of the distribution. However, $\mu_{r,\alpha}(m,t)$ is well defined
for any $m\ge 0$, not necessarily an integer and we will call $\mu_{r,\alpha}(m,t)$ as the generalized $m$-th moment.

Direct computation of $\mu_{r,\alpha}(m,t)$ defined by the rhs of (\ref{mom_ra.1}) from the 
spectral decomposition (\ref{Prxt_sol.2}) is however delicate.
This is because, if we multiply both sides of (\ref{Prxt_sol.2}) by $x^m$ and then integrate over $x$ (after interchanging
the order of integration over $\lambda$ and $x$), we encounter a problem: the integral 
\begin{equation}
\int_0^{\infty} dx\, x^{m+\alpha+1/2}\, J_{-\nu}\left( 2\nu\sqrt{\lambda}\, x^{1/(2\nu)}\right)
\label{non_conv.1}
\end{equation}
is unfortunately not convergent since the Bessel function decays as $x^{-1/2}$ for large $x$ (with oscillations) which
makes the integral divergent. This means that we are not allowed to interchange the order of $\lambda$-integration
and $x$-integration. To circumvent this problem, we use a different trick. Let us first multiply both sides
of Eq. (\ref{Prxt_sol.2}) by the factor $\sqrt{x}\, J_{-\nu}\left(2\nu\sqrt{\lambda'}\, x^{1/(2\nu)}\right)$ and
then integrate over $x\in [0,\infty]$. This makes both the $x$ and the $\lambda$ integration on the rhs
convergent and there is no problem in interchanging the order of integration. This gives
\begin{eqnarray}
&&\int_0^{\infty} P_{r,\alpha}(x,t)\, \sqrt{x}\, J_{-\nu}\left(2\nu\sqrt{\lambda'}\, x^{1/(2\nu)}\right)\, dx
= \nonumber\\
&& \frac{\nu^{1-\nu}}{2\, \Gamma(1-\nu)}\, \int_0^{\infty} d\lambda\, \lambda^{-\nu/2}\, 
M \left( \frac{\lambda}{r+\lambda}, 1, - (r+\lambda) t\right)\, I(\lambda, \lambda')\, ,
\label{mom_ra.2}
\end{eqnarray}
where
\begin{equation}
I(\lambda, \lambda')= \int_0^{\infty} dx\, x^{\alpha+1}\,  J_{-\nu}\left( 2\nu\sqrt{\lambda'}\, x^{1/(2\nu)}\right)\,
 J_{-\nu}\left( 2\nu\sqrt{\lambda}\, x^{1/(2\nu)}\right)\, .
\label{I_def.1}
\end{equation}
Now, we have encountered this integral before in Eq. (\ref{sp_dens.2}). Upon making the change of variable
$2 \nu x^{1/(2\nu)}= y$ and using the identity (\ref{identity_jackson.2}), we simply get
\begin{equation}
I(\lambda,\lambda')= \frac{1}{\nu}\, \delta(\lambda-\lambda')\, .
\label{I_sol.1}
\end{equation}
Substituting this result in (\ref{mom_ra.2}) and carrying out the integral over $\lambda$ gives
\begin{eqnarray}
&&\int_0^{\infty} P_{r,\alpha}(x,t)\, \sqrt{x}\, J_{-\nu}\left(2\nu\sqrt{\lambda'}\, x^{1/(2\nu)}\right)\, dx
= \nonumber\\
&&\frac{\nu^{-\nu}}{2\, \Gamma(1-\nu)}\,  (\lambda')^{-\nu/2}\, 
M \left( \frac{\lambda'}{r+\lambda'}, 1, - (r+\lambda') t\right) \, .
\label{mom_ra.3}
\end{eqnarray}
   
Let us now investigate the left hand side (lhs) of Eq. (\ref{mom_ra.3}). We first use the following series expansion
of Bessel function~\cite{AS_book}
\begin{equation}
J_{-\nu}(z)= 2^{\nu}\, z^{-\nu}\, \sum_{n=0}^{\infty} \frac{(-1)^n\, z^{2n}}{2^{2n}\, n!\, \Gamma(-\nu+n+1)}\, ,
\label{Bessel_series.1}
\end{equation} 
on the lhs of Eq. (\ref{mom_ra.3}) and then integrate term by term of the sum. This gives
\begin{eqnarray}
&&\int_0^{\infty} P_{r,\alpha}(x,t)\, \sqrt{x}\, J_{-\nu}\left(2\nu\sqrt{\lambda'}\, x^{1/(2\nu)}\right)\, dx
= \nonumber\\
&&\nu^{-\nu}\, (\lambda')^{-\nu/2}\, \sum_{n=0}^{\infty} \frac{(-1)^n\, \nu^{2n}}{2\, n!\, \Gamma(-\nu+n+1)}\,
(\lambda')^{n}\, \mu_{r,\alpha}(n/\nu,t)\, ,
\label{lhs_mom.1}
\end{eqnarray}
where the generalized moment $\mu_{r,\alpha}(m,t)$ for general $m\ge 0$ (not necessarily integer) is defined
in Eq. (\ref{mom_ra.1}). Equating this result with the rhs of (\ref{mom_ra.3}) gives
\begin{equation}
\fl
\sum_{n=0}^{\infty} \frac{(-1)^n\, \nu^{2n}}{n!\, \Gamma(-\nu+n+1)}\, (\lambda')^{n}\, \mu_{r,\alpha}(n/\nu,t)=
\frac{1}{\Gamma(1-\nu)}\, M \left( \frac{\lambda'}{r+\lambda'}, 1, - (r+\lambda') t\right) \, .
\label{mom_rel.1}
\end{equation}
Now, the $M$ function on the rhs in (\ref{mom_rel.1}) can, in principle, be expanded in a power series in $\lambda'$ using 
the definition in Eq. (\ref{Kummer_def}) and let us define the coefficients $a_n(t)$ by
\begin{equation}
M \left( \frac{\lambda'}{r+\lambda'}, 1, - (r+\lambda') t\right)= \sum_{n=0}^{\infty} a_n(t)\, (\lambda')^n\, .
\label{M_series.1}
\end{equation}
Substituting this on the rhs of (\ref{mom_rel.1}) and matching the coefficients of the identical powers of $\lambda'$ on both
sides, we get the exact relation 
\begin{equation}
\mu_{r,\alpha}(n/\nu,t)= \frac{(-1)^n\, n!\, \Gamma(n+1-\nu)}{\nu^{2n}\, \Gamma(1-\nu)}\, a_n(t) \, ,
\label{mom_rel.2}
\end{equation}
where $a_n(t)$ is defined in Eq. (\ref{M_series.1}). By using the definition of the Kummer's function (\ref{Kummer_def}) and expanding it at first order in $\lambda'$, one obtains
\begin{equation}
M \left( \frac{\lambda'}{r+\lambda'}, 1, - (r+\lambda') t\right)= 1+\frac{\lambda'}{r}\sum_{m=1}^{\infty}\frac{(-rt)^m}{m\, m!}+{\cal O}(\lambda'^2)\, .
\label{Mexpand}
\end{equation}

As a first check, we deduce $a_0(t)=1$ from Eq. (\ref{Mexpand}) and setting $n=0$ in Eq. (\ref{mom_rel.2}), we get
\begin{equation}
\mu_{r,\alpha}(0,t)= \int_{-\infty}^{\infty} P_{r,\alpha}(x,t)\, dx = 1\, ,
\label{mom0.1}
\end{equation}
that confirms the normalization of the distribution.

A slightly more non-trivial result corresponds to choosing $n=1$ in Eq. (\ref{mom_rel.2}). This gives 
\begin{equation}
\mu_{r,\alpha}\left(\frac{1}{\nu},t\right)= -\frac{(1-\nu)}{\nu^2}\, a_1(t)\, ,
\label{mom_n1.1}
\end{equation}
where $a_1(t)=\frac{1}{r} \sum_{m=1}^{\infty}\frac{(-rt)^m}{m\, m!}$ from the series expansion (\ref{Mexpand}). 
Indeed, the sum appearing in the coefficient $a_1(t)$
was computed explicitly in a similar calculation in Ref. ~\cite{BEM2017},
\begin{equation}
a_1(t)= -\frac{1}{r}\, \left[\ln(rt)+\gamma_E+E_1(rt) \right]\, \quad {\rm where}\quad E_1(z)= 
\int_z^{\infty}\frac{e^{-x}}{x}\, dx\, ,
\label{a1t.1}
\end{equation}
and $\gamma_E=0.5772\ldots$ is the Euler constant.
Substituting $a_1(t)$ in Eq. (\ref{mom_n1.1}) then gives, using $\nu=1/(\alpha+2)$,  the explicit expression valid for any $t$,
\begin{equation}
\mu_{r,\alpha}(\alpha+2,t)=\frac{(\alpha+1)(\alpha+2)}{r}\, \left[\ln(rt)+\gamma_E+E_1(rt)\right]\, .
\label{mom_n1.2}
\end{equation}
\begin{figure}[t]
    \centering
    \includegraphics[width=.5\textwidth,angle=-90]{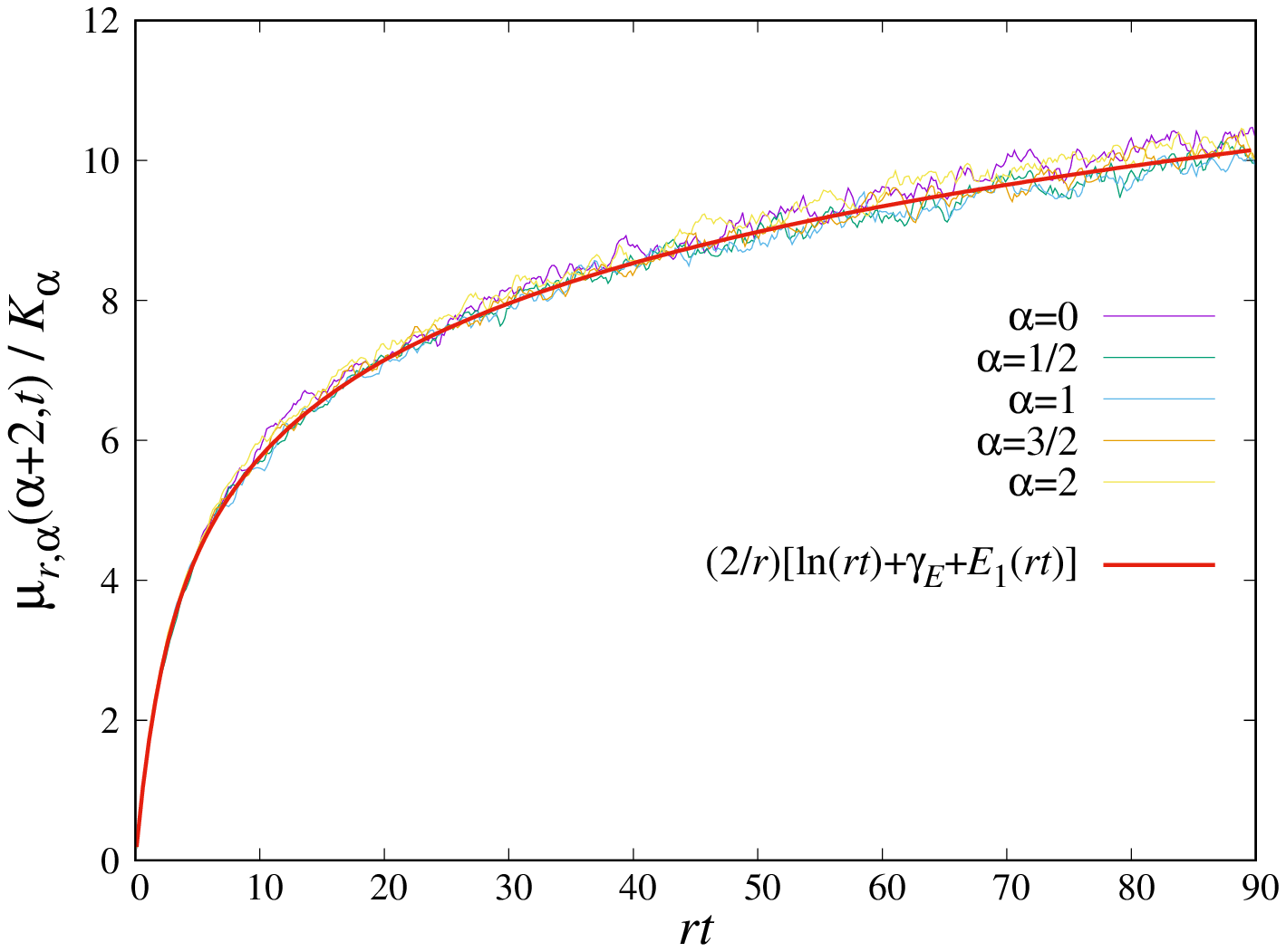}
    \caption{Moment of order $\alpha+2$ rescaled by $K_{\alpha}\equiv (\alpha+1)(\alpha+2)/2$ as a function of time, for various values of $\alpha$. The thin lines correspond to simulations with $r=1$, see  \ref{app:simul} for details. The results are independent of $\alpha$ and follow the evolution predicted by Eq. (\ref{mom_n1.2}) or (\ref{mom_n1.3}).}
    \label{fig:msd}
\end{figure}
Note that for $\alpha=0$, we then recover the result for the variance derived in Ref.~\cite{BEM2017} with $D=1$. 
It is thus convenient to rewrite Eq. (\ref{mom_n1.2}) as
\begin{equation}
\mu_{r,\alpha}(\alpha+2,t)=K_{\alpha}\,\mu_{r,0}(2,t)\, .
\label{mom_n1.3}
\end{equation}
with $K_{\alpha}=(\alpha+1)(\alpha+2)/2$ and where 
\begin{equation}
\mu_{r,0}(2,t)=\frac{2}{r}[\ln(rt)+\gamma_E+E_1(rt)]
\label{msda0}
\end{equation}
is the variance of the memory-walk with $D=1$ \cite{BEM2017}.
Figure \ref{fig:msd} displays $\mu_{r,\alpha}(\alpha+2,t)/K_{\alpha}$ as a function of $rt$ for several values of $\alpha$, where the moments are obtained from numerical simulations of many trajectories following the model rules (see  \ref{app:simul} for a description of the method). As predicted by Eq. (\ref{mom_n1.3}), the different curves collapse onto a single function independent of $\alpha$ and following closely $\mu_{r,0}(2,t)$.

\begin{figure}[t]
    \centering
    \includegraphics[width=.4\textwidth,angle=-90]{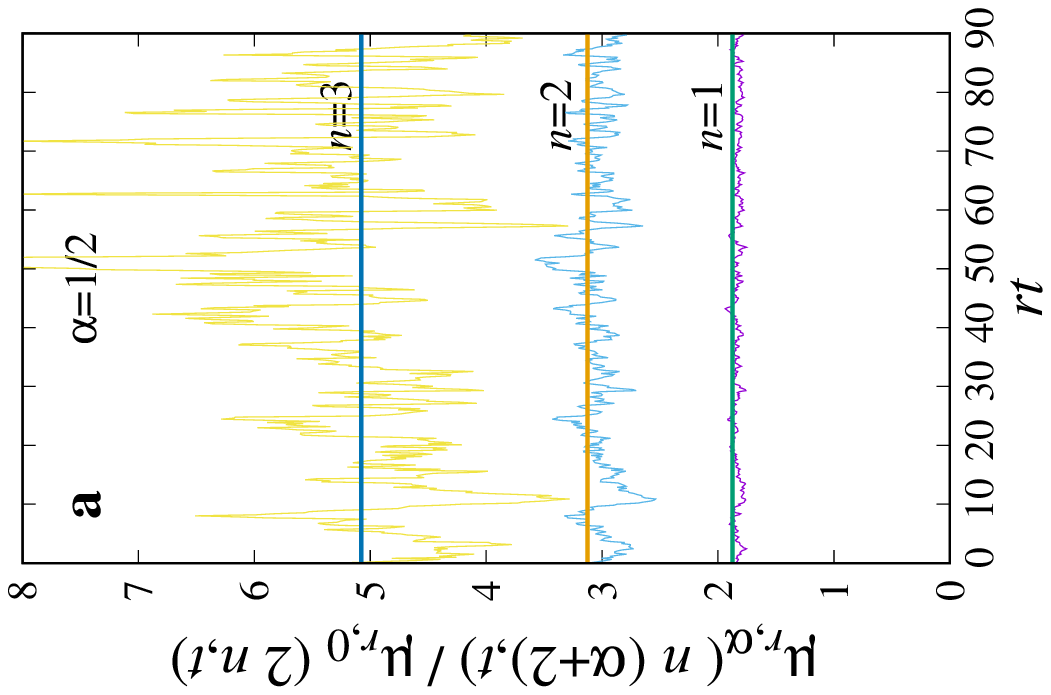}
\includegraphics[width=.4\textwidth,angle=-90]{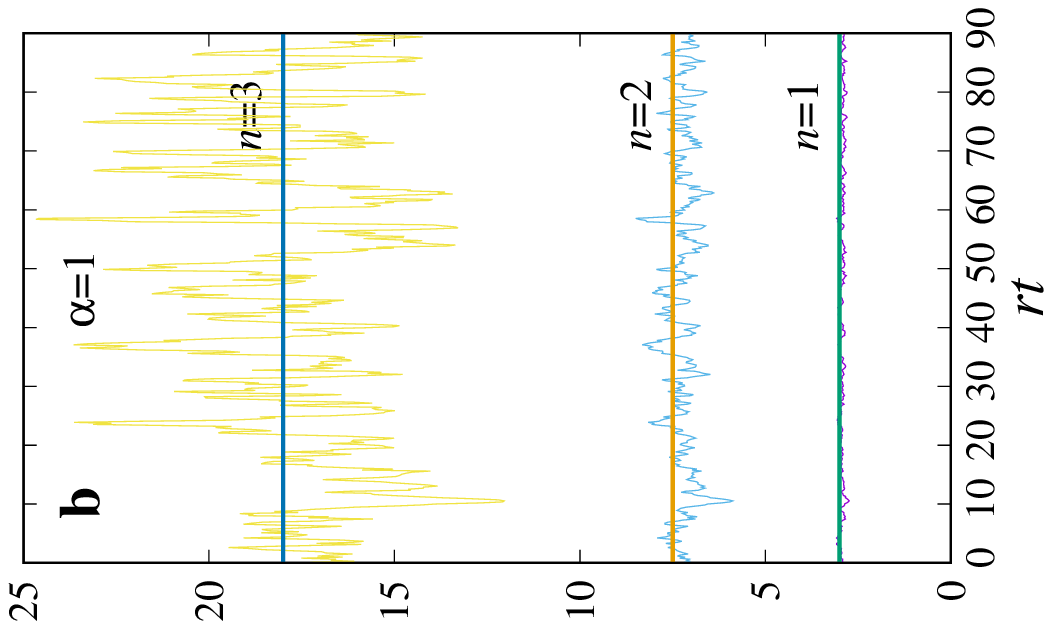}
\includegraphics[width=.4\textwidth,angle=-90]{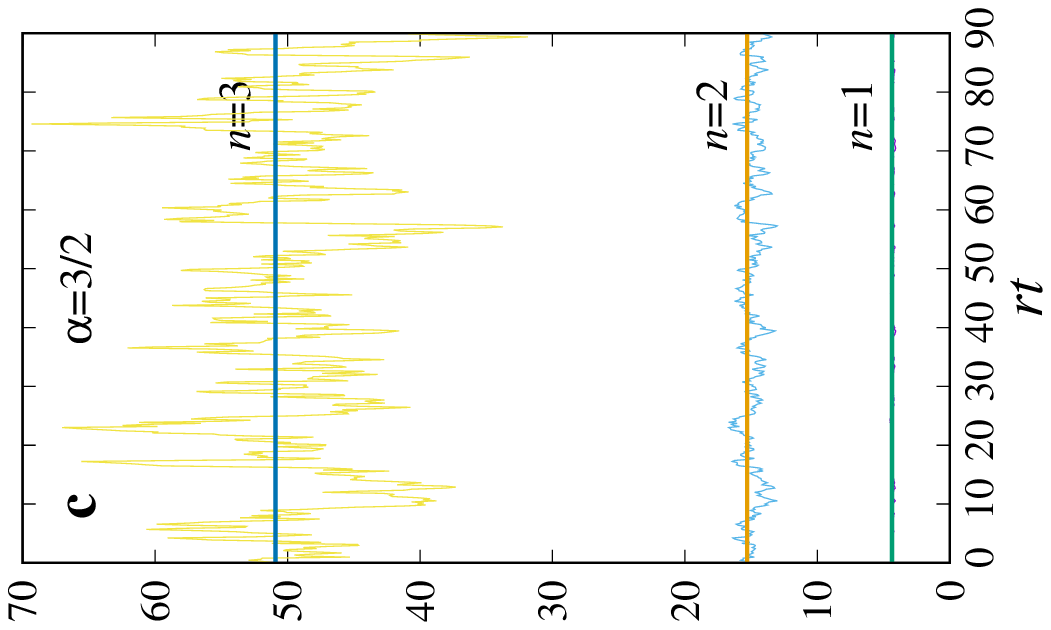}
    \caption{Moment ratio defined by the lhs of Eq. (\ref{int_rel.1}) as a function of $rt$, obtained from simulations with different values of $\alpha$ (a-b-c). The horizontal lines are given by the rhs of Eq. (\ref{int_rel.1}) for $n=1,2,3$. For larger values of $n$, the fluctuations become increasingly visible due to the finite number of realizations ($10^4$). The simulations were performed with $r=1$.}
    \label{fig:comp}
\end{figure}

Finally, let us notice another interesting moment relation in this problem, 
which generalises Eq. (\ref{mom_n1.3}).
Consider the position distribution $P_{r,0}(x,t)$
for $\alpha=0$ given in Eq. (\ref{a0_sol.2}). The Fourier transform of this distribution reads then
\begin{equation}
\tilde{P}_{r,0}(k,t)= \int_{-\infty}^{\infty} P_{r,0}(x,t)\, e^{i\, k\, x}\, dx= M\left(\frac{k^2}{r+k^2}, 1, - (r+k^2)\, t\right)\, .
\label{a0_mom.1}
\end{equation}
Expanding in powers of $k$ on the lhs and rhs (using Eq. (\ref{M_series.1})) gives
\begin{equation}
\mu_{r,0}(2n, t)= (-1)^n\, (2n)!\, a_n(t)\, .
\label{a0_mom.2}
\end{equation}
Eliminating $a_n(t)$ between Eqs. (\ref{mom_rel.2}) and (\ref{a0_mom.2}) provides the relation
\begin{equation}
\frac{\mu_{r,\alpha}(n (\alpha+2), t)}{\mu_{r,0}(2n, t)}= \frac{n!\, \Gamma(n+1-\nu)\, (\alpha+2)^{2n}}{(2n)!\, \Gamma(1-\nu)}\, ,
\label{int_rel.1}
\end{equation}
that is totally independent of time $t$ and the resetting rate $r$. Of course, one can check easily that for $\alpha=0$, the rhs equals unity, as it should. In Figs. \ref{fig:comp}a-b-c, the lhs of Eq. (\ref{int_rel.1}) obtained from numerical simulations is actually consistent with a constant behaviour in time. This ratio fluctuates around a value which is close to the rhs of Eq. (\ref{int_rel.1}). The fluctuations arise from the finite number of trajectories taken to compute the moments and become greater at larger $n$.

\section{Large time behaviour of the variance}\label{sec:msd}
\begin{figure}[t]
    \centering
    \includegraphics[width=.5\textwidth,angle=-90]{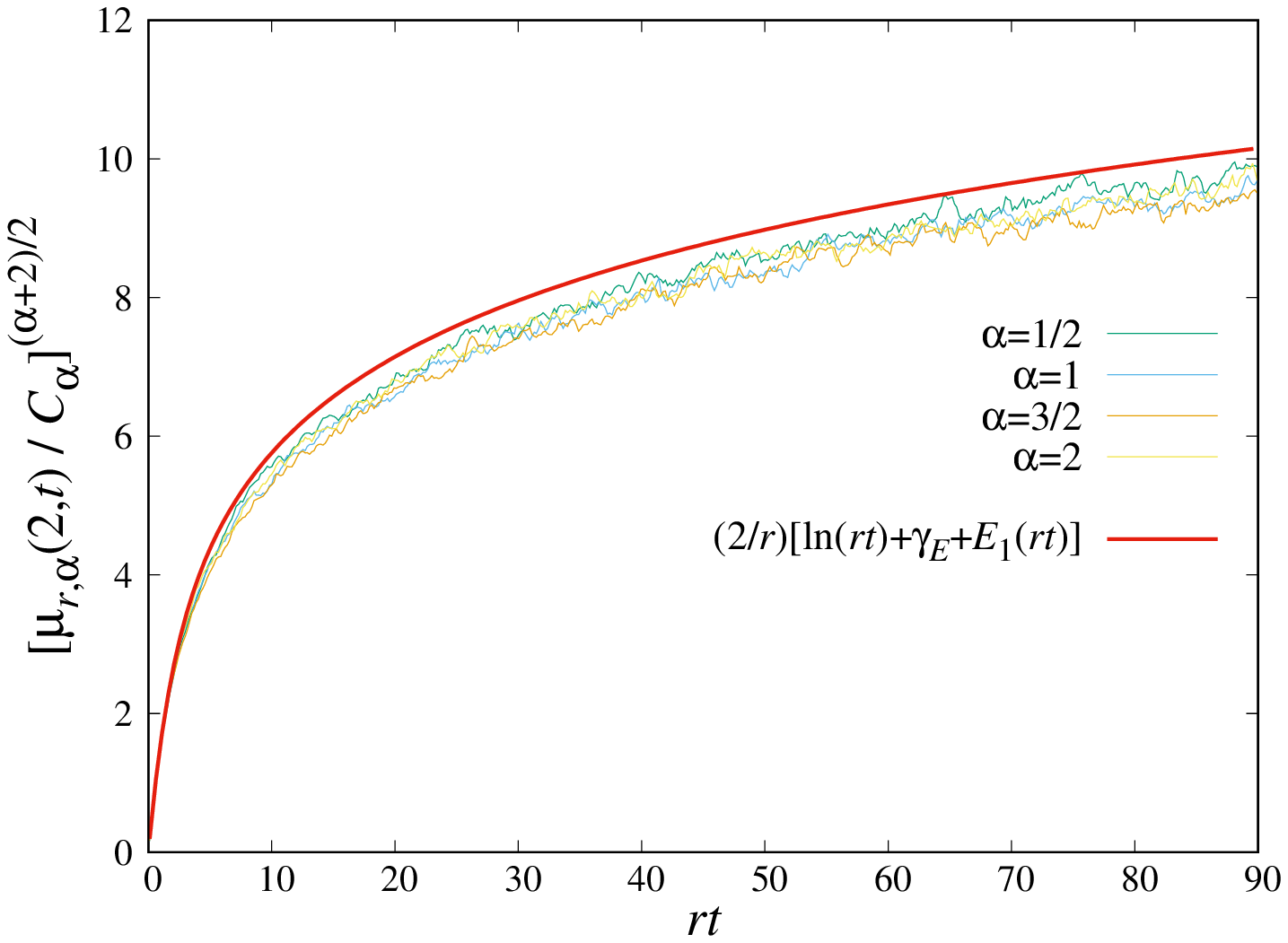}
    \caption{Moment of order $2$ rescaled by $C_{\alpha}$ and taken to the power $(\alpha+2)/2$, as a function of time, for various values of $\alpha$. The thin lines correspond to simulations with $r=1$ and the thick line to the exact moment of order 2 with $\alpha=0$. Over the whole time domain, the numerical results are qualitatively well described by Eq. (\ref{msdar.as2}), which becomes exact when $rt\to\infty$.}
    \label{fig:msd2}
\end{figure}

While the mean square displacement of the particle $\mu_{r,\alpha}(2,t)$ is difficult to obtain exactly for all $t$ if $r\neq 0$ and $\alpha\neq 0$, its asymptotic behaviour can be derived from the scaling law (\ref{Prxt_asymp.1}). At very large times, the latter relation becomes increasingly accurate and
\begin{equation}
\mu_{r,\alpha}(2,t)\approx 2\xi(t)^2\int_0^{\infty}dz\,z^2G_{\alpha}(z)\,,\quad{\rm with}\ 
\xi(t)=\left(\frac{\ln (rt)}{r}\right)^{\nu}\, ,
\end{equation}
where $G_{\alpha}(z)$ is given by Eq. (\ref{scaling.1}). After integration, 
\begin{equation}
\mu_{r,\alpha}(2,t)\approx \nu^{-4\nu} \frac{\Gamma(1+\nu)}{\Gamma(1-\nu)}\,
\left( \frac{\ln (rt)}{r} \right)^{2/(\alpha+2)}\,.
\label{msdra.as1}
\end{equation}
We can proceed a bit further by noticing from Eq. (\ref{msda0}) that $\ln(rt)/r$ and $\mu_{r,0}(2,t)/2$ tend to be equal at late times, i.e., $\lim_{t\to\infty}[\ln(rt)/r]/[\mu_{r,0}(2,t)/2]=1$.  Therefore, at leading order, Eq. (\ref{msdra.as1}) can be recast as
\begin{equation}
\mu_{r,\alpha}(2,t)\approx C_{\alpha} [\mu_{r,0}(2,t)]^{2/(\alpha+2)}\,,
\label{msdar.as2}
\end{equation}
with 
\begin{equation}
C_{\alpha}=2^{-2\nu}\nu^{-4\nu}\frac{\Gamma(1+\nu)}{\Gamma(1-\nu)}\,.
\label{ca}
\end{equation}
The form (\ref{msdar.as2}) makes the relationship between the two moments more explicit, and
unlike Eq. (\ref{mom_n1.3}), it is not exact at all $t$ (except for the trivial case $\alpha=0$, where it is easy to check that $C_0=1$). We can nevertheless test the accuracy of Eq. (\ref{msdar.as2}) at short and intermediate $t$, where $\mu_{r,0}(2,t)$ is exactly known in Eq. (\ref{msda0}). Figure \ref{fig:msd2} displays the time dependence of $[\mu_{r,\alpha}(2,t)/C_{\alpha}]^{(\alpha+2)/2}$ obtained from numerical simulations.  The curves depend little on $\alpha$ and turn out to be quite close to the expression of $\mu_{r,0}(2,t)$ in the time interval $0\le rt\le 90$. Therefore Eq. (\ref{msdar.as2}) provides a reasonable description of the variance over the whole time domain. Notice that $rt=90$ cannot be considered as belonging to the scaling regime, since its logarithm is only $\simeq4.49$.  

\section{Generalization to higher dimensions}
\label{higher_d}

Our main result for the position distribution $P_r(x,t)$ in Eq. (\ref{Prxt_sol.2}) for one dimension can be easily extended to higher
dimensions $d\ge 1$. In $d\ge 1$, we choose $D(\vec x)= D(R)$ where $R=|\vec x|$ in order to preserve the spherical
symmetry. The associated Fokker-Planck equation for the position distribution
$P_r(\vec x, t)$ reads
\begin{equation}
\partial_t P_r(\vec x,t)= \nabla^2 \left[D(R)\, P_r(\vec x,t)\right] - r\, P_r(\vec x,t) +
\frac{r}{t}\, \int_0^t dt' \,  P_r(\vec x, t')\,  ,
\label{fpr_d.1}
\end{equation}
starting from the initial condition
\begin{equation}
P_r(\vec x, t=0)= \delta(\vec x)\, .
\label{init_d.1}
\end{equation}
Since $D(R)$ depends only on the radial distance $R$ from the centre, the solution $P_r(\vec x, t)= P_r(R,t)$
also depends only on the radial distance. Using this spherical symmetry, Eq. (\ref{fpr_d.1}) then reduces to
\begin{eqnarray}
\partial_t P_r(R,t)&=& \left[ \frac{{\partial }^2}{\partial R^2} + \frac{(d-1)}{R}\, \frac{\partial}{\partial R}\right]\left[D(R)\, P_r(R,t)\right] - r\, P_r(R,t)\nonumber\\ 
&&+ \frac{r}{t}\, \int_0^t {\rm d}t' \,  P_r(R, t')\,  .
\label{fpr_d.2}
\end{eqnarray}
The method of separation of variables works exactly as in $d=1$. Writing, $P_r(R,t)= f_r(t)\, \phi(R)$, one finds that
the time-dependent part $f_r(t)$ is exactly the same as in Eq. (\ref{ft_sol.1}) for $d=1$, namely
\begin{equation}
f_r(t)=M \left( \frac{\lambda}{r+\lambda}, 1, - (r+\lambda) t\right)\, .
\label{ft_d.1}
\end{equation}
where $M(a,b,z)$ is the Kummer's function. We now choose $D(R)=R^{-\alpha}$ The space-dependent part $\phi(R)$ depends on $d$ as well as $\lambda$.
As in $d=1$, we define 
\begin{equation}
h_{\lambda}(R)= \frac{\phi_{\lambda}(R)}{R^{\alpha}}\, ,
\label{h_def}
\end{equation}
which then satisfies the ordinary differential equation
\begin{equation}
\frac{{\rm d}^2 h_{\lambda}(R)}{{\rm d}R^2}+ \frac{(d-1)}{R}\,\frac{{\rm d} h_\lambda(R)}{{\rm d}R}+  \lambda\, R^{\alpha}\, h_{\lambda}(R)=0\, .
\label{hR_diff.1}
\end{equation}
Note that the function $h_\lambda(R)$ implicitly depends on $d$, but for notational simplicity we do not display this explicitly.

The differential equation (\ref{hR_diff.1}) can be solved exactly and the most general solution is given by
\begin{eqnarray}
&&h_{\lambda}(R)= B_1(\lambda)\, R^{(2-d)/2}\, J_{(d-2)\,\nu}\left( 2\, \nu\, \sqrt{\lambda}\, R^{1/(2\nu)}\right)\nonumber\\
&& +B_2(\lambda)\, R^{(2-d)/2}\, \ J_{-(d-2) \nu}\left(2\, \nu\, \sqrt{\lambda}\, R^{1/(2\nu)}\right)\, \quad {\rm where}\ \nu= 
\frac{1}{\alpha+2}\, ,
\label{hd_sol.1}
\end{eqnarray}
where $B_1(\lambda)$ and $B_2(\lambda)$ are two unknown constants (independent of $R$). 
Once again, the constants $B_1(\lambda)$ and $B_2(\lambda)$ implicitly depends on $d$, but we do not display
this explicitly for simplicity.
Using the small $z$ behaviour
of $J_\nu(z)$ in Eq. (\ref{Jz_asymp.1}), one finds that while the term multiplying $B_1(\lambda)$ in Eq. (\ref{hd_sol.1})
is perfectly analytic as $R\to 0$, the second term multiplying $B_2(\lambda)$ is non-analytic as $R\to 0$, and behaves as
$R^{2-d}$. Thus the second derivative of $h_\lambda(R)$ diverges as $R^{-d}$ as $R\to 0$, if
$B_2(\lambda)$ is non-zero. Since the second derivative must remain finite
as $R\to 0$ [the basic assumption that leads to the differential equation (\ref{hR_diff.1})], we must have
$B_2(\lambda)=0$. Taking the product of the space and time-dependent parts and integrating over all possible $\lambda\ge 0$,
we then arrive at the exact spectral decomposition
\begin{eqnarray}
\fl
P_{r,\alpha}(R,t)= &&\\
 &&\fl R^{\alpha+(2-d)/2}\, \int_0^{\infty} d\lambda\, 
B_1(\lambda)\, J_{(d-2)\nu}\left( 2\, \nu\, \sqrt{\lambda}\, R^{1/(2\nu)}\right)\,
M \left( \frac{\lambda}{r+\lambda}, 1, - (r+\lambda) t\right)\,  ,\nonumber
\label{Prxt_sol_d.1}
\end{eqnarray}
where we have restored the dependence on $\alpha$ explicitly in the position distribution by denoting it
by $P_{r,\alpha}(R,t)$.

The spectral density $B_1(\lambda)$ is again determined from the initial condition (\ref{init_d.1}) as in $d=1$. Setting $t=0$ in
Eq. (\ref{Prxt_sol_d.1}) and using $M(a, b, 0)=1$ we get
\begin{equation}
R^{\alpha+(2-d)/2}\, \int_0^{\infty} d\lambda\, B_1(\lambda)\, J_{(d-2)\nu}\left( 2\, \nu\, \sqrt{\lambda}\, R^{1/(2\nu)}\right)=
\delta(\vec x)\, .
\label{sp_dens_d.1}
\end{equation}
We now multiply both sides by $R^{d/2}\, J_{(d-2)\nu}\left( 2\,\nu\, \sqrt{\lambda'}\, R^{1/(2\nu)}\right)$ 
and integrate over $R\in [0,\infty]$. This gives
\begin{eqnarray}
&&\fl \int_0^{\infty} d\lambda\, B_1(\lambda)\, \int_0^{\infty} dR\, R^{\alpha+1}\, 
J_{(d-2)\nu}\left( 2\,\nu\, \sqrt{\lambda}\, R^{1/(2\nu)}\right)\,
J_{(d-2)\nu}\left( 2\,\nu\, \sqrt{\lambda'}\, R^{1/(2\nu)}\right) \nonumber\\
&&\fl =\int \delta(\vec x)\, R^{d/2}\, J_{(d-2)\nu}\left( 2\, \nu\, \sqrt{\lambda}\, R^{1/(2\nu)}\right)\, dR\, .
\label{decomp.1}
\end{eqnarray}
The main contribution to the integral on the rhs of (\ref{decomp.1})
comes from the $R\to 0$ region. Using the small argument expansion (\ref{Jz_asymp.1})
we get
\begin{equation}
{\rm rhs} = \frac{\nu^{(d-2)\nu}\, \left(\lambda'\right)^{(d-2)\nu/2}}{\Gamma\left(1+(d-2)\nu\right)}\, \int \delta(\vec x)\, R^{d-1}\, dR\, .
\label{rhs.2}
\end{equation}
However, the integral $\int \delta(\vec x)\, R^{d-1}\, dR= 1/\Gamma_d$ where $\Gamma_d= 2\, \pi^{d/2}/\Gamma(d/2)$ is the
volume of a $d$-dimensional unit sphere. Hence, we finally get the rhs as
\begin{equation}
{\rm rhs} = \frac{\Gamma(d/2)}{2\, \pi^{d/2}}\, \frac{\nu^{(d-2)\nu}\, \left(\lambda'\right)^{(d-2)\nu/2} }{\Gamma\left(1+(d-2)\nu\right)}\, .
\label{rhs.3}
\end{equation}
Similarly, the lhs of Eq. (\ref{decomp.1}) can be performed exactly, as in $d=1$, by first making the change of variable
$2\,\nu\, R^{1/(2\nu)}=y$ and then using the identity
(\ref{identity_jackson.2}). This finally determines $B_1(\lambda)$ explicitly in arbitrary dimension
\begin{equation}
B_1(\lambda)= \frac{\Gamma(d/2)}{2\, \pi^{d/2}}\, \frac{ \nu^{(d-2)\nu+1}\, 
\lambda^{(d-2)\nu/2} }{\Gamma\left(1+(d-2)\nu\right)}\, .
\label{B1d.1}
\end{equation}
Finally, putting this solution for $B_1(\lambda)$ back to Eq. (\ref{Prxt_sol_d.1}), we obtain our exact position distribution
in $d$ dimension
\begin{eqnarray}
&&P_{r,\alpha}(R,t)= \frac{\Gamma(d/2)}{2\, \pi^{d/2}}\, \frac{ \nu^{(d-2)\nu+1}}{\Gamma\left(1+(d-2)\nu\right)}\,
R^{\alpha+ (2-d)/2}\, \nonumber\\
&&\times\int_0^{\infty} d\lambda\,  \lambda^{(d-2)\nu/2}\,J_{(d-2)\nu}\left( 2\, \nu\, \sqrt{\lambda}\, R^{1/(2\nu)}\right)\,
M \left( \frac{\lambda}{r+\lambda}, 1, - (r+\lambda) t\right)\, , 
\label{Prxt_sol_d_final}
\end{eqnarray}
where we recall that $\nu=\frac{1}{\alpha+2}$.
It is easy to check that for $d=1$, we recover the solution (\ref{Prxt_sol.2}).

As in $d=1$, one can derive the scaling behaviour of $P_r(R,t)$ in Eq. (\ref{Prxt_sol_d_final}) 
at late times. Following exactly as in $d=1$ (and omitting
the steps since they are very similar), we find that
in the limit $t\to \infty$, $R\to \infty$, but with the ratio $z=R\left( \frac{r}{\ln (rt)}\right)^{\nu}$ fixed, the
position distribution $P_r(R,t)$ in $d$ dimension approaches the scaling form
\begin{equation}
P_{r,\alpha}(R,t) \approx 
\left(\frac{r}{\ln (rt)}\right)^{\nu\, d}\, G_{\alpha,d}\left(\frac{r^{\nu}}{\left(\ln (rt)\right)^{\nu}}\, R\right)\, 
\label{Pr_scaling_d.1}
\end{equation}
where the scaling function $G_{\alpha,d}(z)$ is given by the explicit form
\begin{equation}
\fl
G_{\alpha,d}(z)= \frac{\Gamma(d/2)}{2\, \pi^{d/2}}\, \frac{ \nu^{(d-2)\nu+1}}{\Gamma\left(1+(d-2)\nu\right)}\,
z^{\alpha}\, \exp\left[- \nu^2\, z^{\alpha+2}\right]\, \quad {\rm with}\quad \nu=\frac{1}{\alpha+2}\, .
\label{scaling_function_d}
\end{equation}
Clearly, for $d=1$, this scaling function reduces to $G_{\alpha,1}(z)= G_{\alpha}(z)$ given in Eq. (\ref{scaling.1}).
Interestingly, the scaling function $G_{\alpha, d}(z)$ depends on $d$ only 
through the overall normalizing factor, but the
$z$-dependence of $G_{\alpha,d}(z)$ is independent of $d$.

\section{Summary and Conclusion}
\label{summary}

We have studied a sluggish random walk model consisting of a Brownian particle starting from 
the origin and whose diffusion coefficient decays algebraically with the distance from this 
point, or $D(x)=|x|^{-\alpha}$.  In addition, the walker uses from time to time (with a rate 
$r$) a memory rule whereby it resets to positions previously visited in the past. These 
revisits are preferential, i.e., the probability to reset to a small region of space is 
proportional to the total amount of time spent there up to the present time. This model can 
be considered as a simple animal movement model, where an individual is reluctant to move 
away from a central place and at the same time exploits known places in its environment.

We have obtained the exact solution of the position distribution of the particle $x(t)$ at 
all times in arbitrary spatial dimensions and have focused our analysis on the one-dimensional 
case, which already captures the main properties. While the model in the absence of memory ($r=0$) 
exhibits a subdiffusive behaviour, with $x(t)\sim t^{1/(\alpha+2)}$ typically, the combined 
effects of the vanishing diffusion coefficient at large distances and of the returns to 
frequently visited positions lead to an extremely slow, sub-logarithmic diffusion at late 
times, as $x(t)\sim [\ln(rt)]^{1/(\alpha+2)}$. Despite this sluggishness, the walker does 
not become localized in space asymptotically: its position distribution keeps depending on 
time and tends to a scaling law. The associated scaling function is the same as in the 
memory-less case with $r=0$. However, contrary to this latter problem, when $r>0$ scaling 
does not hold at all times but only asymptotically.  Due to the form of $D(x)$, the particle 
is repelled from the origin, hence the scaling function is a non-monotonous function of the 
distance.

Obtaining the moments of the position, such as the mean square displacement, at all times is 
rather difficult even in one dimension. However, rather interestingly, it is 
possible to compute at all times the moment of order $\alpha+2$, i.e., $\langle 
|x|^{\alpha+2}\rangle$ explicitly in $d=1$. It closely resembles at all times the mean 
square displacement of the problem with $\alpha=0$ (a diffusion coefficient constant in 
space), to which it is simply related by a proportionality factor. Similarly, the moments of 
order $n(\alpha+2)$ are proportional at all time to those of order $2n$ of the problem with 
$\alpha=0$. These analytical results in $d=1$ have been verified via numerical simulations.

It would be interesting to study the localized steady states and localization dynamics of 
the model presented here in the presence of a specific resource site, thus extending 
previous work with a diffusion coefficient constant in space \cite{FCBGM2017,BFGM2019,MB2025}.

\vspace{0.5cm}
{\bf Acknowledgements:} DB acknowledges support from CONACYT (Mexico) Grant
CF2019/10872 and from the University of Paris-Saclay, France (scientific missionary).
SNM acknowledges support from ANR Grant No. ANR-23-CE30-0020-01 EDIPS.

\appendix 

\section{Numerical Simulations}\label{app:simul}
\label{numerical}

We have generated stochastic trajectories by implementing the following numerical scheme (see also \cite{BM2024_1}). Time is discrete, $t=n\Delta t$ with $n=0,1,2,\ldots$ and $\Delta t\ll 1$ a fixed time-step. As $n$ increases by one unit, with probability $1-r\Delta t$, the position $x(n\Delta t)\equiv X_n$ of the particle is updated via the discretized version of the Langevin equation (\ref{lange.1}) in one dimension,
\begin{equation}
X_{n+1}=X_n+\sqrt{2D(X_n)\Delta t}\,\xi_{n+1}\, ,
\end{equation}
where $\{\xi_{n}\}_{n=1,2,\ldots}$ are i.i.d. Gaussian random variables with zero mean and unit variance. To avoid a diverging diffusivity near $x=0$ and therefore very large increments, we choose
\begin{equation}
D(x)=\frac{1}{|x|^{\alpha}+\epsilon^{\alpha}}\, ,
\end{equation}
with $\epsilon$ a crossover length such that $\epsilon\ll 1$ and $\epsilon^{-\alpha}\Delta t\ll 1$. For practical purpose $\Delta t$ is fixed to $10^{-5}$, and we choose $\epsilon=10^{-2}$ if $1\le\alpha\le 2$, while $\epsilon=10^{-4}$ if $\alpha=1/2$. We expect the long term dynamics to be independent of the choice of $\epsilon$ as soon as it fulfils the above conditions. 
Averages are performed over $10^4$ or $6\times 10^4$ independent simulations, for $r\neq0$ and $r=0$, respectively.

With the complementary probability $r\Delta t$, the next position is determined by the memory rule: a time $t'$ is chosen uniformly in the interval $[0,t]$, i.e., an integer $n'$ is chosen uniformly among $\{0,1,2,\ldots,n\}$, and the walker is reset to the position it occupied at that  time in the past,
\begin{equation}
X_{n+1}=X_{n'}\,.
\end{equation}
To obtain the numerical scaling functions of Fig. \ref{fig:scal} with $r=0$, the simulation time $t$ was set to $10$.

\end{document}